\documentclass[a4paper,11pt]{article}

\usepackage{pos}
\usepackage{lipsum}
\usepackage[left]{lineno}

\title{Prospects for GeV Neutrino Transient Searches with the IceCube Upgrade}

\ShortTitle{GeV Neutrino Transient Searches with IceCube Upgrade}

\author{The IceCube Collaboration \\{\normalsize \normalfont(a complete list of authors can be found at the end of the proceedings)}\\}

\emailAdd{kobayashi@hepburn.s.chiba-u.ac.jp}

\abstract{

The recent detection of TeV neutrino sources by the IceCube Neutrino Observatory demonstrates the detector's advanced capabilities in detecting high-energy astrophysical neutrinos. At lower energies, down to the GeV range, a variety of transient phenomena, such as novae, supernovae, and gamma-ray bursts, are expected to emit neutrinos. Observations of these neutrinos can provide unique insights into processes below the photosphere and offer clues to identifying their emission mechanisms. We have searched for these neutrinos intensively with IceCube's existing infill array, DeepCore. Although no significant detections have been made, strong constraints on astrophysical environments in these transients, such as the baryon loading factor in gamma-ray bursts, have been obtained. A denser infill array, called the IceCube Upgrade, will enhance sub-TeV neutrino searches with its unprecedented sensitivity to GeV neutrinos. The Upgrade, set to be deployed in the 2025-2026 South Pole season, will consist of seven new strings, adding approximately 700 novel optical modules with multiple photomultiplier tubes in the DeepCore volume. The denser arrangement of high-efficiency modules will significantly improve IceCube's sensitivity between 1 GeV and 1 TeV. We present an initial assessment of the astrophysical capabilities of the IceCube Upgrade, using preliminary simulated data and an event selection similar to that used for DeepCore. We explore the detectability of GeV neutrino transients compared to DeepCore and discuss potential sensitivity enhancements through advanced detector simulations and optimized analysis techniques, including refined triggering conditions and event selection criteria.

\vspace{4mm}

{\bfseries Corresponding authors:}
Yukiho Kobayashi$^{1*}$\\
{$^{1}$ \itshape The International Center for Hadron Astrophysics, Chiba University}\\
$^*$ Presenter
}

\ConferenceLogo{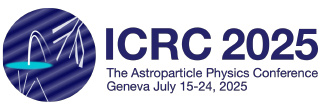}

\FullConference{39th International Cosmic Ray Conference (ICRC2025)\\
 15–24 July 2025\\
Geneva, Switzerland\\}

\begin{document}

\maketitle

\section{Introduction}

Thanks to their straight-line propagation and {low interaction cross-section with matter}, neutrinos serve as powerful messengers for probing extreme environments in astrophysical sources.
They also play a unique role in identifying cosmic-ray accelerators in the Universe.
A wide variety of astrophysical transients, such as gamma-ray bursts (GRBs), supernovae (SNe), and novae, are expected to emit neutrinos, with GeV neutrinos potentially carrying crucial information about the environments of these sources~\cite{Partenheimer:2024qxw}.
Searching for these neutrinos is therefore essential for understanding the nature of these transients.

GRBs are among the most intriguing astrophysical transients.
They are the most energetic explosions in the Universe, and their powerful jets may accelerate cosmic rays to ultra-high energies $\sim10^{20}\rm~eV$~\cite{grb_uhecr1994}.
At GeV energies, GRBs may emit quasithermal neutrinos, which are produced through inelastic collisions between protons and neutrons after they have decoupled from each other in the relativistic outflow~\cite{Murase2013}.
Importantly, these neutrinos are naturally predicted without relying on any cosmic-ray acceleration mechanism in shocks.
Since quasithermal neutrinos originate below the photosphere, their detection would support the photospheric scenario of the GRB prompt emission~\cite{PhysRevD.78.101302}, whose mechanism is a long-standing question in high-energy astrophysics.

The typical energy of quasithermal neutrinos is estimated to be $E_\nu\approx30\left(\frac{\Gamma}{100}\right)\left(\frac{\Gamma_{\rm rel}}{3}\right)~\rm GeV$, where $\Gamma$ is the bulk Lorentz factor of the GRB jets and $\Gamma_{\rm rel}$ is the relative Lorentz factor between the protons and neutrons~\cite{Murase2013}.
Thus, GRBs with relatively low Lorentz factors, or low-luminosity GRBs (LLGRBs), are promising sources of quasithermal neutrinos in the GeV energy range.
LLGRBs are thought to be failed GRBs, whose jets are choked by stellar matter and do not fully escape the progenitor.
Detection of quasithermal neutrinos from LLGRBs can thus provide valuable insights into the GRB-SN connection.
LLGRBs are also promising sources for the observed astrophysical diffuse neutrino flux in the TeV-PeV energy range, which makes them even more intriguing~\cite{llgrb_to_diffuse_nu2016}.
Therefore, detecting GeV neutrinos from GRBs is crucial for a comprehensive understanding of this class of transients.

Another promising class of sources for quasithermal neutrino production at GeV energies is core-collapse SNe.
If a rapidly rotating and magnetized protoneutron star (PNS) is left as a remnant after a SN, it can launch a relativistic outflow~\cite{Murase2014, Carpio2024}.
As in the GRB case, neutrons are initially accelerated together with ions, but decouple as the density decreases during expansion.
These neutrons eventually collide with the decelerated ejecta, producing quasithermal neutrino in the 0.1-10 GeV energy range.
Observations of these neutrinos would offer unique insights into the jet formation processes in SN and help to establish a unified understanding of the GRB and SN mechanisms.
Detecting GeV neutrinos also plays a key role in bridging the thermal emission at MeV energies and non-thermal emission at TeV energies~\cite{Abe_2021, Murase:2017_SNTeVnu}.

Novae are another promising source of GeV neutrinos.
About 20 novae have been detected in GeV gamma rays by Fermi-LAT so far~\cite{Chomiuk:2020zek}.
Among them, the 2021 outburst of RS Ophiuchi was the first to be detected in very-high-energy gamma rays by imaging atmospheric Cherenkov telescopes~\cite{MAGIC:2022rmr, HESS:2022qap, CTA-LSTProject:2025kbv}.
These gamma-ray observations suggest cosmic-ray acceleration in the nova shocks, which is expected to result in accompanying neutrino emission in a comparable energy range.
Detecting such neutrinos would further support the hadronic origin of the gamma-ray emission from novae~\cite{jessie:2025icrc}.

The IceCube Neutrino Observatory is a Cherenkov detector that instruments $1~\rm km^3$ of ice at the South Pole with 5,160 optical sensors buried at depths of up to 2,500 meters.
The densely instrumented infill array in the middle of the IceCube fiducial volume, DeepCore, is designed to lower the energy threshold.
Dedicated event selections based on DeepCore, GeV Reconstructed Events with Containment for Oscillation (GRECO)~\cite{Abbasi_2022, IceCube:2022lnv} and the Extremely Low-Energy (ELOWEN) sample~\cite{2021PhRvD.103j2001A}, have been developed and applied to searches for various transients in the energy ranges of 10-1000 GeV and 0.5-5 GeV, respectively.
Although no significant detection of transient sources has been made so far~\cite{greco_gw_2023, IceCube:2022lnv}, these analyses have placed strong constraints on the physical properties of these transients, such as the baryon loading factor of GRB jets~\cite{IceCube:2023woj}.

The IceCube Upgrade is a low-energy extension of the existing DeepCore detector and is scheduled for deployment during the 2025-2026 austral summer~\cite{Ishihara:2019aao}.
The Upgrade will consist of seven new strings installed in the middle of the DeepCore fiducial volume, each instrumented with 100 novel optical modules.
The majority of the Upgrade optical modules, called D-Eggs and mDOMs, are equipped with multiple photomultiplier tubes per module, significantly enhancing their effective photodetection area with respect to the original IceCube modules~\cite{IceCube:2022mng, Mechbal:2023BO}.
These Upgrade modules are even more densely deployed than DeepCore, with approximately 3 meters of vertical and 20-30 meters of horizontal spacing, which is expected to lead to significant improvement in sensitivity at GeV energies.

\section{Simulation and Reconstruction}

This work is based on the same simulation and reconstruction used in studies of the Upgrade's sensitivity to atmospheric neutrino oscillations~\cite{IceCube:2023ins}.
Neutrino interactions are simulated using GENIE version 2.12.8, while atmospheric muons are generated with MuonGun, a toolkit for efficient muon simulation based on the technique introduced in Ref.~\cite{BECHERINI20061}.
The injected neutrino energy range is 1 GeV to 500 GeV.
The propagation of muons is treated with PROPOSAL~\cite{KOEHNE20132070}, and the other particles are propagated and decayed with 
GEANT4~\cite{AGOSTINELLI2003250}.
Cherenkov photons emitted in ice are propagated using CLSim with the latest IceCube ice model~\cite{clsim, icemodel2024}.
The detector response is simulated to convert the detected photons into a series of readout pulses, individual signal outputs from the photosensors, characterized by their charge and time.
Noise pulses, arising from sources such as radioactive decay in the vessel glass of the optical modules, are also added.
An event is defined based on a trigger that requires a certain number of coincident pulses in both space and time.

The reconstruction of event kinematics, i.e., the neutrino energy and direction, with an associated uncertainty on the direction, is performed using graph convolutional neural networks (GNNs).
The GNN models are trained and applied using GraphNet, an open-source deep learning library for neutrino telescopes~\cite{Søgaard2023}.
A GNN model is also employed to remove pulses likely originating from noise prior to the reconstruction.

Further event selection is applied to suppress background from atmospheric muons and noise.
The selection process includes a veto filter to reject incoming muons, simple straight cuts based on event quantities, such as the number of pulses and the number of hit modules, and classifications based on machine-learning techniques.
The event selection adopted here is comparable to the ``IC93'' sample described in Ref.~\cite{IceCube:2023ins}.
This event sample, hereafter referred to as IC93, is dominated by atmospheric neutrinos with a rate of 6.6~mHz, and contains $\sim2\rm~mHz$ noise events and 0.8~mHz atmospheric muons, resulting in a total rate of 9.5~mHz.

\begin{figure}
    \centering
    \includegraphics[width=0.75\linewidth]{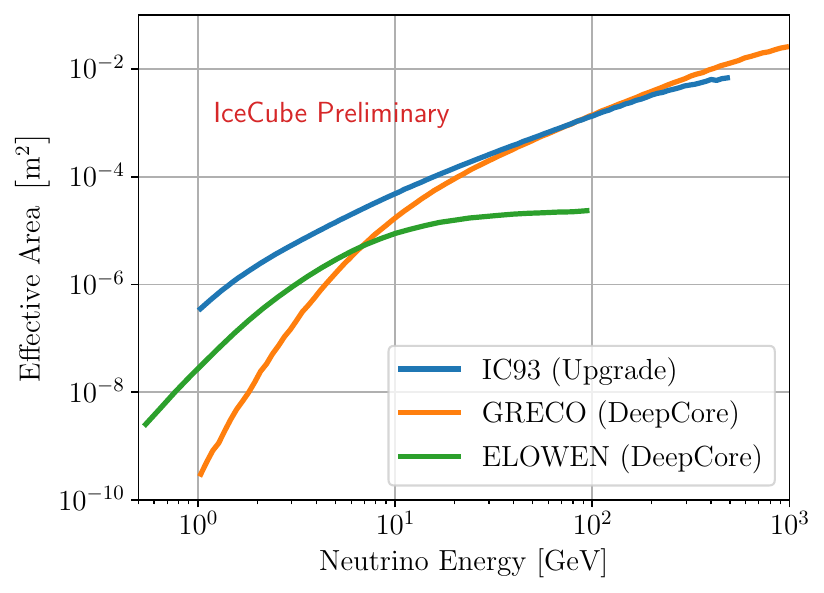}
    \caption{Sky-averaged, all-flavor summed, and $\nu+\overline{\nu}$ averaged effective area of the IC93 sample in comparison to the existing DeepCore event selections.
    The ELOWEN effective area is that of Ref.~\cite{icrcube_icrc2025_first_elowen}.}
    \label{fig:Aeff}
\end{figure}

Figure~\ref{fig:Aeff} shows the all-flavor summed and $\nu+\overline{\nu}$ averaged effective area of the IC93 sample in comparison to those of the existing DeepCore event selections.
One can see the Upgrade significantly enhances the effective area at energies below 30 GeV, which is essential to improve sensitivity to transients in this energy range.

\section{Analysis}

Mock background data are constructed by sampling the simulated neutrinos and muons according to their atmospheric distributions.
Although noise events are not yet included in the mock data, this is not expected to change the final results drastically, given that the noise events would increase the total background event rate by only $\sim30\%$.
Moreover, the noise events are expected to be further reduced by onboard noise suppression on mDOMs, which is not considered in the current simulation. 

In this work, we evaluate the sensitivity of the IceCube Upgrade to transients, assuming the source is point-like and that both its spatial coordinates and burst time window are known.
The sensitivity is estimated with an unbinned maximum likelihood method, which is widely employed in neutrino transient searches~\cite{BRAUN2008299}.
An extended likelihood approach is adopted to account for variations in the total number of observed events, which is particularly relevant for short-duration observations where event statistics are limited.
Given a probability density function for signal, $\mathcal{S}$, and a background PDF, $\mathcal{B}$, the extended likelihood is defined as
\begin{equation}
    \mathcal{L}(n_s, \gamma) = \frac{(n_s + n_b)^N e^{-(n_s+n_b)}}{N!} \times \prod_{i=1}^{N} \left[ \frac{n_s}{n_s + n_b} \mathcal{S}(x_i | \gamma) + \frac{n_b}{n_s + n_b} \mathcal{B}(x_i) \right],
\label{eq:likelihood}
\end{equation}
where $N$ is the number of events in the sample, $n_s$ and $n_b$ are the expected number of signal and background events, respectively, and $\gamma$ is the spectral index of the source.
The index $i$ runs over all observed events, and each $x_i$ represents the set of observables for a given event, namely its reconstructed energy and direction.

The source spatial PDF is modeled using a Kent function, with a width according to the per-event angular uncertainty, while the background spatial PDF is directly parameterized from the simulated event sample.
Both the signal and background energy PDFs are also parameterized from the simulated sample, weighted to the respective spectra.
The differential sensitivity is evaluated in half-decade energy bins, assuming a power-law source spectrum, $F=\Phi_0\left(E/E_0\right)^{-\gamma}$, with spectral index of $\gamma=2$ in each energy bin.
For calculating sensitivity to specific theoretical models, which are shown in Figure \ref{fig:diff_sens}, their spectral shapes are approximated with a power-law spectrum with an exponential cutoff, $F=\Phi_0\left(E/E_0\right)^{-\gamma}\exp{\left(-E/E_{\rm cut}\right)}$.
For a model of neutrino emission from a nova, the $\nu_\mu + \overline{\nu_\mu}$ flux expectation in Ref.~\cite{MAGIC:2022rmr}, based on 2021 gamma-ray observations of RS Ophiuchi, is used.

When analyzing a single transient source, the test statistic (TS) for an observed event sample is calculated as the log-likelihood ratio between the best-fit hypothesis, realized at $n_s =\hat{n}_s$, and the background-only hypothesis, where $n_s=0$:
\begin{equation}
    \mathcal{T} \mathcal{S} = -2 \hat{n}_s + 2 \sum_{i=1}^{N} \ln \left[ \frac{\hat{n}_s}{n_b}\frac{\mathcal{S}(x_i)}{\mathcal{B}(x_i)} + 1 \right].
    \label{eq:ts}
\end{equation}
The sensitivity is then defined as the flux that results in a TS greater than the median TS of background-only trials in 90\% of cases, corresponding to a 90\% confidence level (CL) upper limit (UL).

\section{Results}

\begin{figure}
    \centering
    \includegraphics[width=0.75\linewidth]{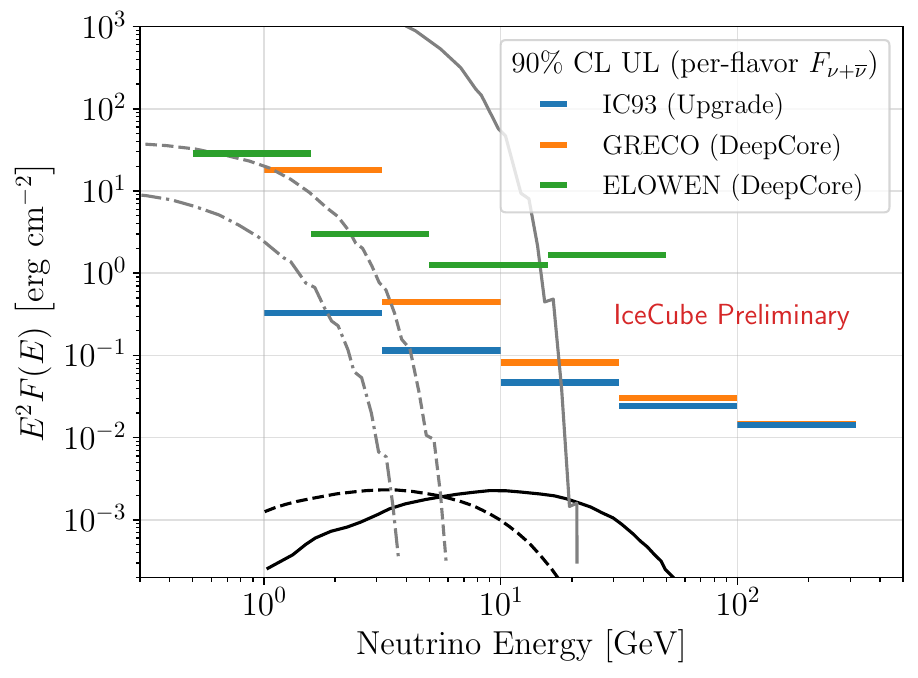}
    \caption{Differential sensitivity of the IceCube Upgrade compared to the existing DeepCore event selections, under observational conditions of a time window width $\Delta T = 1000~\rm s$ and a source declination of $20^\circ$.
    The ELOWEN sensitivity is taken from Ref.~\cite{IceCube:icrc2023boat}.
    The black curves show $\nu_\mu + \bar{\nu}_\mu$ emission models for LLGRBs located at 10~Mpc, with bulk Lorentz factors of $\Gamma = 30$ (solid line) and $\Gamma = 10$ (dashed line), from~Ref.~\cite{Murase2013}.
    The gray curves represent per-flavor neutrino emission models from a core-collapse SN at 10~kpc, forming a rapidly rotating and strongly magnetized PNS with $B = 10^{15}\rm~T$.
    The solid, dashed, and dash-dotted lines correspond to rotation periods of $P = 1$, 3, and 5~ms, respectively, from Ref.~\cite{Carpio2024}.}
    \label{fig:diff_sens}
\end{figure}

The estimated differential sensitivity of the IceCube Upgrade to transient sources is shown in Figure~\ref{fig:diff_sens}, in comparison with the existing DeepCore event selections.
This evaluation assumes an observational time window of $\Delta T = 1000\rm~s$ and a source declination of $\delta = 20^\circ$.
This particular declination is chosen to enable a fair comparison with the ELOWEN sensitivity, taken from Ref.~\cite{IceCube:icrc2023boat}.
As seen in the figure, the Upgrade significantly enhances the sensitivity, especially at energies $E \lesssim 10~\rm GeV$.
For time windows of $\Delta T \lesssim 1000~\rm s$, point-source searches with the Upgrade are effectively background-free, so the 90\% CL UL corresponds to $n_s = 2$–$3$ events.
This is also true for GRECO, and the improvement with the Upgrade can be primarily attributed to its larger effective area, as shown in Figure~\ref{fig:Aeff}.
The lower background rate in the IC93 sample ($\lesssim9\rm~mHz$) than that in ELOWEN ($\sim20\rm~mHz$)~\cite{2021PhRvD.103j2001A}, along with the enhanced effective area, accounts for the improvement over ELOWEN.
In addition, the effective background rate in the IC93 sample is further reduced thanks to its angular resolution.
ELOWEN does not have direction reconstruction.

\begin{figure}
    \centering
    \includegraphics[width=1\linewidth]{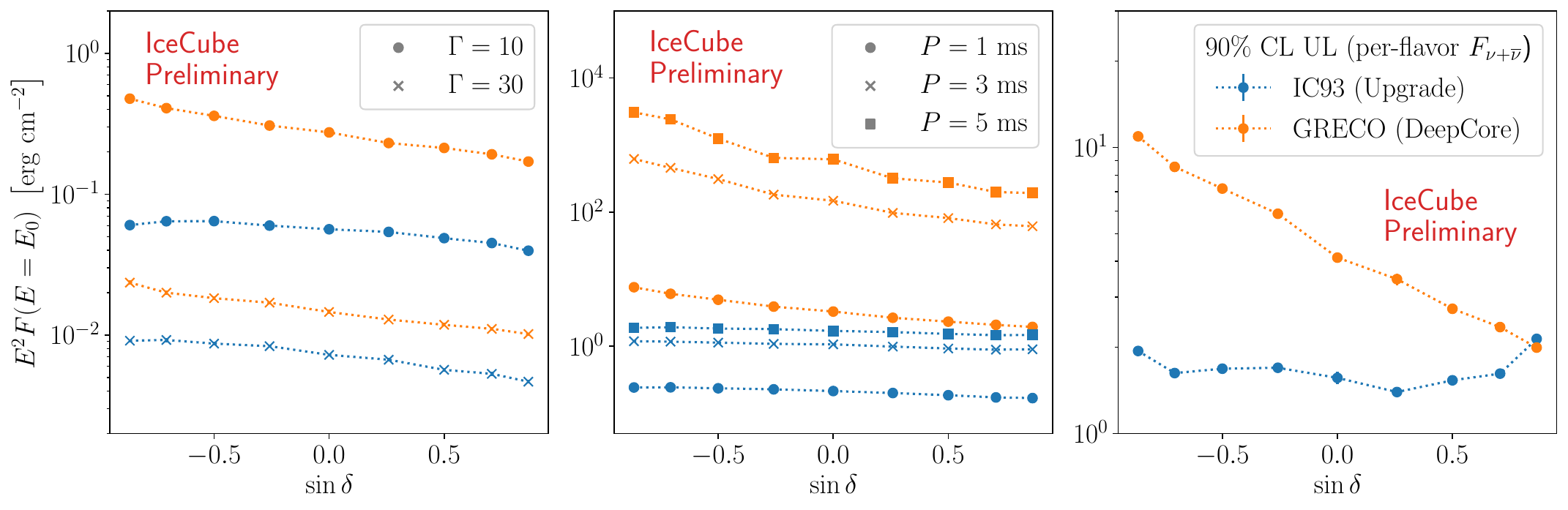}
    \caption{Sensitivity of the IceCube Upgrade (blue) and GRECO (orange) to theoretical neutrino emission models as a function of the source declination.
    The flux is normalized at $E_0=1\rm~GeV$.
    Left: Sensitivity to LLGRBs with the bulk Lorentz factor $\Gamma$, assuming a time window of $\Delta T=1000\rm~s$.
    Middle: Sensitivity to core-collapse SNe forming a PNS with a magnetic field of $B=10^{15}~\rm T$ and a rotation period of $P$, assuming $\Delta T=\rm100~s$.
    Right: Sensitivity to novae, assuming $\Delta T=\rm~10$~days.
    }
    \label{fig:sens_to_model}
\end{figure}

Estimated sensitivities to specific emission models are presented in Figure~\ref{fig:sens_to_model}.
In the left panel, the Upgrade shows a factor of 2–5 improvement in sensitivity to LLGRBs compared to GRECO, depending on the GRB's bulk Lorentz factor.
This translates to a detectable distance of approximately $2~\rm Mpc$.
Thus, LLGRBs occurring in nearby galaxies may be detectable with the Upgrade, although such events are expected to be extremely rare, given the estimated LLGRB rate of $10^{-7}$–$10^{-6}~\rm Mpc^{-3}yr^{-1}$\cite{Liang_2007, Sun_2015}.

The middle panel of Figure~\ref{fig:sens_to_model}, when compared with the SN emission models shown in Figure~\ref{fig:diff_sens}, indicates that the Upgrade is capable of detecting neutrinos from a Galactic SN forming a rapidly rotating and strongly magnetized ($B = 10^{15}\rm~T$) PNS.
The Upgrade is sensitive to the models with rotation periods up to $P = 3$–$5\rm~ms$, which would be difficult to detect with DeepCore alone.
Moreover, extreme PNS models (e.g., $B = 10^{15}\rm~T$ and $P = 1\rm~ms$) could be constrained even if observed in nearby galaxies at Mpc-scale distances.

The right panel of Figure~\ref{fig:sens_to_model} shows that sensitivity to novae is also improved with the Upgrade, particularly in the southern sky.
This is beneficial since novae are Galactic sources and are predominantly distributed in the southern hemisphere.
Although detecting a single nova remains challenging, a nearby and exceptionally bright nova would offer a unique opportunity to place stringent constraints on neutrino emission from novae or detect a signal.
One promising candidate is T Coronae Borealis, which is predicted to undergo an outburst in the next few years~\cite{2023MNRAS.524.3146S, jessie:2025icrc}.

\section{Conclusion and Outlook}

The upcoming IceCube Upgrade, scheduled for deployment during the 2025–2026 austral summer, is anticipated to significantly enhance the capabilities of IceCube for detecting astrophysical transients in the GeV energy range.
In this work, we evaluate the Upgrade's sensitivity to transient sources using an unbinned maximum likelihood method, based on preliminary simulation and event selection.
We find that the Upgrade offers substantial improvements in sensitivity below 100 GeV, particularly below 10~GeV, compared to the existing DeepCore event selections.
This enhancement increases the prospects for detecting GeV transients.
For example, the Upgrade can detect Galactic SNe forming a rotating and magnetized PNS, even under moderate model assumptions.
LLGRBs may also be detectable if they occur in nearby galaxies.
The sensitivity to novae is also expected to improve significantly.

Both the simulation and event selection used in this study are still preliminary, leaving room for further improvement.
For instance, relaxing the trigger conditions will enhance signal acceptance, which is critical for transient searches.
This may be feasible, especially considering the expected capabilities of the mDOM onboard noise suppression.
Additional optimization may also be possible in the event classification by machine learning techniques.
The IC93 sample evaluated here strongly suppresses muons using machine-learning classifiers optimized for neutrino oscillation studies.
For transient searches at GeV energies, however, muon rejection does not have to be so powerful.
Looser selections could improve sensitivity to GeV transients.
Further improvements might also be achieved through more advanced detector simulations, for instance, by incorporating the angular distribution of Cherenkov photons at the modules, and by developing reconstruction algorithms that take advantage of this information.

\footnotesize
\bibliographystyle{ICRC}
\bibliography{references}

\clearpage

\section*{Full Author List: IceCube Collaboration}

\scriptsize
\noindent
R. Abbasi$^{16}$,
M. Ackermann$^{63}$,
J. Adams$^{17}$,
S. K. Agarwalla$^{39,\: {\rm a}}$,
J. A. Aguilar$^{10}$,
M. Ahlers$^{21}$,
J.M. Alameddine$^{22}$,
S. Ali$^{35}$,
N. M. Amin$^{43}$,
K. Andeen$^{41}$
% ...
\\
\\
% ...
$^{10}$ Universit{\'e} Libre de Bruxelles, Science Faculty CP230, B-1050 Brussels, Belgium \\
% ...
$^{16}$ Department of Physics, Loyola University Chicago, Chicago, IL 60660, USA \\
$^{17}$ Dept. of Physics and Astronomy, University of Canterbury, Private Bag 4800, Christchurch, New Zealand \\
% ...
$^{21}$ Niels Bohr Institute, University of Copenhagen, DK-2100 Copenhagen, Denmark \\
$^{22}$ Dept. of Physics, TU Dortmund University, D-44221 Dortmund, Germany \\
% ...
$^{35}$ Dept. of Physics and Astronomy, University of Kansas, Lawrence, KS 66045, USA \\
% ...
$^{39}$ Dept. of Physics and Wisconsin IceCube Particle Astrophysics Center, University of Wisconsin{\textemdash}Madison, Madison, WI 53706, USA \\
% ...
$^{41}$ Department of Physics, Marquette University, Milwaukee, WI 53201, USA \\
% ...
$^{43}$ Bartol Research Institute and Dept. of Physics and Astronomy, University of Delaware, Newark, DE 19716, USA \\
% ...
$^{63}$ Deutsches Elektronen-Synchrotron DESY, Platanenallee 6, D-15738 Zeuthen, Germany \\
$^{\rm a}$ also at Institute of Physics, Sachivalaya Marg, Sainik School Post, Bhubaneswar 751005, India \\
% ...

\subsection*{Acknowledgments}

\noindent
The authors gratefully acknowledge the support from the following agencies and institutions:
USA {\textendash} U.S. National Science Foundation-Office of Polar Programs,
U.S. National Science Foundation-Physics Division,
U.S. National Science Foundation-EPSCoR,
U.S. National Science Foundation-Office of Advanced Cyberinfrastructure,
Wisconsin Alumni Research Foundation,
Center for High Throughput Computing (CHTC) at the University of Wisconsin{\textendash}Madison,
Open Science Grid (OSG),
Partnership to Advance Throughput Computing (PATh),
Advanced Cyberinfrastructure Coordination Ecosystem: Services {\&} Support (ACCESS),
Frontera and Ranch computing project at the Texas Advanced Computing Center,
U.S. Department of Energy-National Energy Research Scientific Computing Center,
Particle astrophysics research computing center at the University of Maryland,
Institute for Cyber-Enabled Research at Michigan State University,
Astroparticle physics computational facility at Marquette University,
NVIDIA Corporation,
and Google Cloud Platform;
Belgium {\textendash} Funds for Scientific Research (FRS-FNRS and FWO),
FWO Odysseus and Big Science programmes,
and Belgian Federal Science Policy Office (Belspo);
Germany {\textendash} Bundesministerium f{\"u}r Forschung, Technologie und Raumfahrt (BMFTR),
Deutsche Forschungsgemeinschaft (DFG),
Helmholtz Alliance for Astroparticle Physics (HAP),
Initiative and Networking Fund of the Helmholtz Association,
Deutsches Elektronen Synchrotron (DESY),
and High Performance Computing cluster of the RWTH Aachen;
Sweden {\textendash} Swedish Research Council,
Swedish Polar Research Secretariat,
Swedish National Infrastructure for Computing (SNIC),
and Knut and Alice Wallenberg Foundation;
European Union {\textendash} EGI Advanced Computing for research;
Australia {\textendash} Australian Research Council;
Canada {\textendash} Natural Sciences and Engineering Research Council of Canada,
Calcul Qu{\'e}bec, Compute Ontario, Canada Foundation for Innovation, WestGrid, and Digital Research Alliance of Canada;
Denmark {\textendash} Villum Fonden, Carlsberg Foundation, and European Commission;
New Zealand {\textendash} Marsden Fund;
Japan {\textendash} Japan Society for Promotion of Science (JSPS)
and Institute for Global Prominent Research (IGPR) of Chiba University;
Korea {\textendash} National Research Foundation of Korea (NRF);
Switzerland {\textendash} Swiss National Science Foundation (SNSF).

\end{document}